\documentclass{article}
\usepackage{mrs2005,epsfig}
\begin{document} 
\title{THE VERY LATE-STAGE MERGER IC 4200: PRE-ELLIPTICAL OR PRE-SPIRAL?}

\author{Paolo Serra$^1$, S.C. Trager$^1$, J.M. van der Hulst$^1$, T.A. Oosterloo$^{1,2}$, R. Morganti$^{1,2}$, J.H. van Gorkom$^{1,3}$}
\affil{$^1$Kapteyn Astronomical Institute, $^2$ASTRON, $^3$Columbia University}

\begin{abstract}

Recent observations have revealed a class of unusually HI-rich early-type galaxies. By combining observations of their morphology, stellar populations and neutral hydrogen we aim to understand how these galaxies fit into the hierarchical formation paradigm. Here we present the result of our radio and optical observations of a test case galaxy, the E/S0 IC 4200.
\end{abstract} 
 
\section{Introduction}

Observations during the last decades have been changing the traditional picture of early-type galaxies. Once thought to be old, passively evolving and kinematically-relaxed systems devoid of gas, early-type galaxies have been discovered to be a class with subtle but frequent (hence important) peculiarities: among these, HI structures surrounding the stellar body. The hierarchical scenario provides an interpretation of the these observations. According to it, the gas in a peculiar early-type galaxy is the result of recent accretion of a gas-rich satellite galaxy. The external origin is suggested by the fact that, as opposed to the situation in spirals, the HI content does not correlate with the properties of the underlying stellar body. However, recent observations have found galaxies with extremely massive, extended and quite regular HI structures, difficult to explain in terms of satellite accretion. The natural question to ask is whether these objects fit in the hierarchical scenario.

Barnes (2002) showed that such extended gas disks could be formed during the merger of gas-rich galaxies as a result of the re-accretion of gas that conserved its angular momentum, forming the tidal bridges and tails observed in many merging systems, and later settled into an extended but dilute rotationally supported disk. On the other hand, Keres et al.\ (2005) showed that IGM gas can be accreted around galaxies via a cold mode; this could result in atomic gas structures around galaxies.

If mergers are responsible for the formation of the observed gas systems, other signatures of the recent formation event should be found in the host galaxies, such as star formation occurring at the time of the merger as a consequence of gas infall and morphological fine structures indicating the incomplete relaxation of the stellar body. To understand if hierarchical assembly plays the dominant role, we have started an observational project aimed to study the HI gas content, stellar populations and optical morpholgy of HI-rich early-type galaxies. Here we present our study of the test case galaxy IC 4200, an E/S0 at z=0.013 and with $L_B$=5.30$\times{10^{10}}$L$_{\odot}$.

\section{Radio and Optical Observations}

Following-up the HIPASS detection, we observed IC 4200 with the ATCA (Australian Telescope Compact Array) to characterise its HI-gas properties. We found a $\sim{90}$ deg warped HI disk of $8.54\times{10^9}$ M$_{\odot}$ extended out to 60 kpc from the center of the galaxy (Fig.1); the gas of the disk is very dilute and as a result, no large-scale detectable star formation is hosted. A faint plume of HI reveals a gas flow from IC 4200 to an HI-rich companion 110 kpc distant. The two galaxies do not show any hint of interaction at optical wavelengths. The morphology and velocity field of the HI around IC 4200 are regular and symmetric around the center within a radius of 30 kpc. Assuming the rotational velocity of 340 km/s that we find from modelling the HI disk, the gas must have been in place for at least 1-2 Gyr to settle into a smooth configuration at this radius.

%  
% Figure 1 
% 
\begin{figure}
\vspace*{1.25cm}  
\begin{center}
\epsfig{figure=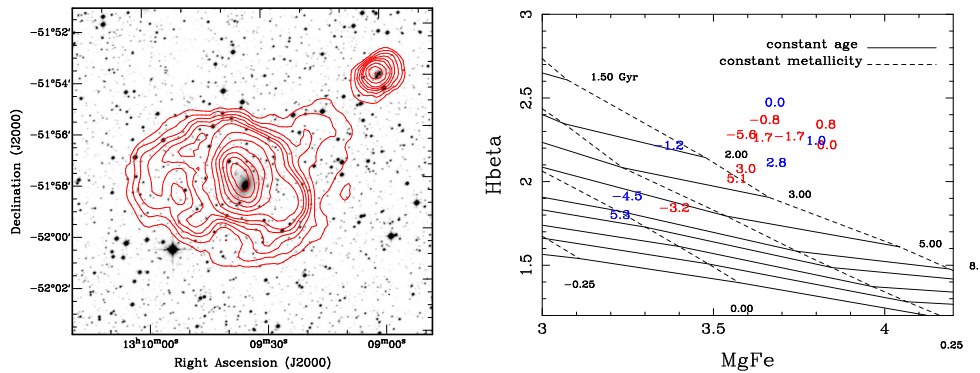,width=13cm}  
\end{center}
\vspace*{0.25cm}
\caption{\small{\emph{Left:} Total HI contours over an optical image of IC4200; contours levels are (0.24, 0.40, 0.57, 0.74, 0.91, 1.08, 1.41, 2.09, 2.42, 2.76, 3.09, 3.36)x1020 cm-2. 8.5$\times{10^9}$ and 7.3$\times{10^8}$ M$_{\odot}$ of HI were detected around IC 4200 and its companion respectively. \emph{Right:} IC 4200 line indices MgFe and H$\beta$ along the major (red) and minor (blue) axis plotted over Worthey Ô94 models grids. Labels indicate the distance from the center of the galaxy in arcsec.}} 
\end{figure}

To study the stellar phase of IC 4200, we obtained high-S/N two-dimensional long-slit optical spectroscopy as well as $V$- and $R$-band imaging at ESO/La Silla Observatory using NTT/EMMI. Spectroscopy revealed ubiquitous presence of ionized gas at any distance from the center, with ionization typical of LINER's. This indicates that some activity is taking place within the stellar body. Confirming this, 1.4 GHz continuum emission is ten times the value expected from 60 $\mu$m emission in the case of star formation. Along the major axis, we detected stellar rotation and, in the central region, gas counter-rotation. Along the minor axis, only gas rotation is detected. Altogether, ionized gas rotates independently from stars and in a way consistent with the larger scale kinematics of the HI disk, suggesting that ionized and atomic gas might share a common history.

For stellar populations analysis we first extracted one-dimensional bins from the two-dimensional spectrum to sample radial gradients. We removed the ionized gas emission from each of them by fitting Bruzual \& Charlot (2003) stellar populations models, by building a model gas emission by fitting each feature of the residual with a gaussian and finally by removing the model gas emission from the original spectrum. We measured Lick/IDS absorption-line-strength indices and compared them to Worthey (1994) models extended to non-solar abundance ratios (Trager et al.\ 2000). We found SSP-equivalent ages of $\sim{2}$ Gyr in the center of the galaxy (Fig.1).

Finally, we fitted the $V$- and $R$-band images with Sersic and Sersic $+$ exponential profiles to find morphological fine structures. Independently of the fitting function, we found shells aligned along the major axis.

\section{Discussion}

Scaling Barnes' (2002) results to the stellar mass of IC 4200, we find that two Milky Ways with twice the HI abundance of our Galaxy can produce, within 1 Gyr from the merger, a warped gas disk whose mass and extension match the observed one. Combining this scenario with the indications provided by the kinematics and morphology of the disk, it is possible that IC 4200 formed via a merger more than 1-2 Gyr ago. Furthermore, the stellar populations age can be explained if a burst of star formation was induced by a merger occurred less than 2 Gyr ago. Finally, shells aligned to the major axis can be formed as a result of a major merger (Hernquist \& Spergel 1992). Putting these three observational results together, it is possible that IC 4200 formed via a major merger 2 Gyr ago.

Based on the results of Keres et al.\ (2005), it is also possible that the observed HI disk formed via cold accretion from the IGM. However, we calculated that in this case the bulk of mass accretion must have occurred before $z$=1, while no substantial accretion took place in the last $\sim{6}$ Gyr. It is not clear how this could match the indications coming from stellar populations. Furthermore, IGM accretion fails to explain the formation of shells. Therefore, we propose that IC 4200 was formed via a merger of two gas-rich galaxies $\sim{2}$ Gyr ago.

Although we found that IC 4200 is the result of hierarchical assembling, it is unlikely that it will further evolve into a normal early-type galaxy. The observed gas disk, too dilute to host star formation, might stay intact for very long times. Another possibility is that continuing gas infall increases the gas density enough to initiate star formation and make IC 4200 become a spiral. Therefore, it seems that hierarchical formation events can led a wider range of morphologies than simple early-type galaxies.

\vfill 
\end{document}